\documentclass[12pt,timesmath]{article}
\usepackage{amssymb}
\usepackage{amsmath}
\newtheorem{ass}{General Assumption}
\newtheorem{theorem}[ass]{Theorem}
\newtheorem{lemma}[ass]{Lemma}

\newtheorem{corollary}[ass]{Corollary}
\begin{document}

\title{On the stability of the Kerr metric}

\author{Horst R. Beyer \\
 Max Planck Institute for Gravitational Physics, \\
 Albert Einstein Institute, \\
 D-14476 Golm, Germany}

\date{\today}                                     

\maketitle

\begin{abstract}
The reduced (in the angular coordinate $\varphi$) 
wave equation and Klein-Gordon equation 
are considered on a Kerr background and in the 
framework of $C^{0}$-semigroup theory. Each equation
is shown to have a well-posed initial value problem,
i.e., to have a  unique solution depending 
continuously on the data. Further, it is shown that the 
spectrum of the semigroup's generator coincides with the 
spectrum of an operator polynomial whose coefficients can 
be read off from the equation. In this way the problem of 
deciding stability is reduced to a spectral problem and
a mathematical basis is provided for mode considerations.
For the wave equation it is shown that the resolvent of 
the semigroup's generator and the corresponding Green's 
functions can be computed using spheroidal functions. 
It is to be expected that,
analogous to the case of a Schwarzschild background,  
the quasinormal frequencies of the Kerr black hole 
appear as {\it resonances}, i.e., poles of the analytic continuation 
of this resolvent. 
Finally, stability of the background with respect to 
reduced massive perturbations is proven for large enough masses. 
\end{abstract}


\section{Introduction}

The stability of the Schwarzschild black hole was demonstrated 
by Kay and Wald \cite{kaywald} who showed the boundedness of 
all solutions of the wave equation corresponding to $C^{\infty}$ 
data of compact support. Their proof rests on the positivity of 
the conserved energy.
\newline  
\linebreak 
The problem is more subtle for Kerr space time.
A conserved energy exists, but the energy
density is negative inside the ergosphere. 
Hence the total energy could be finite while the 
field still might grow exponentially in parts of the spacetime. 
Papers by Press and Teukolsky \cite{pressteuk}, Hartle and Wilkins
\cite{hartlewilkins}, and Stewart \cite{stewart} make the
absence of exponentially growing normal modes very 
plausible. Whiting \cite{whiting} has proven that 
there are no such modes, and in his proof he showed
that normal modes grow at most linearly in time. 
Recent numerical evolution calculations \cite{krivan1,krivan2} 
for slowly and fast rotating Kerr black holes show no sign of 
exponential growth. In the case of massive scalar 
perturbations of Kerr results of Damour,Deruelle and Ruffini \cite{damour},
Zouros and Eardley \cite{zouros}, and Detweiler   
\cite{damour} point to the existence of unstable modes.
These modes are very slowly growing 
with growth times similar to the age of universe.
This fact complicates the numerical detection 
of such modes. 
\newline
\linebreak
Here we consider the reduced (in the angular coordinate $\varphi$) 
wave equation and Klein-Gordon equation on a Kerr background and in the 
framework of $C^{0}$-semigroup theory. For this the 
mathematical framework from \cite{beyer} 
is used. 
For each equation 
it is shown that the initial value problem is
well-posed, i.e., has a unique solution which depends
continuously on the data. Further, it is shown that the 
spectrum of the semigroup's generator coincides with the 
spectrum of an operator polynomial whose coefficients can 
be read off from the equation. In this way the problem of 
deciding stability is reduced to a spectral problem.
For the wave equation it is shown that the resolvent of 
the semigroup's generator and the corresponding Green's 
functions can be computed using spheroidal functions. 
It is to be expected that, analogous to the case of a 
Schwarzschild background, the quasinormal frequencies of the Kerr black hole 
appear as poles of the analytic continuation of this resolvent. 
Finally, the stability of the background with respect to 
reduced massive perturbations is proven for large enough masses.
This is done by applying an abstract stability criterium from \cite{beyer}.
\newline 
\linebreak
The Kerr metric in Boyer-Lindquist coordinates $t,r,\theta,\varphi$
is given by
\begin{eqnarray}
g = &&\left(1 - \frac{2Mr}{\Sigma}\right)dt^2 + 
\frac{4Mar\sin^2\theta}{\Sigma}dtd\varphi -
\frac{\Sigma}{\triangle}dr^2 -
\Sigma d\theta^2  \nonumber \\
&& - \left(r^2 + a^2 + \frac{2Ma^2r\sin^2}{\Sigma}\right) \sin\theta d\varphi^2
\, \, ,
\end{eqnarray}
where $M$ is the mass, $a \in [0,M]$ is the rotational
parameter,
\begin{equation}
\triangle := r^2 - 2Mr +a^2 \, \, , \, \, \Sigma := r^2 + a^2 \cos^2 \theta 
\, \, .
\end{equation}
The coordinates are constrained by 
$ - \infty < t < + \infty$ , $r_{+} < r < + \infty$ , $-\pi < \varphi < \pi$
and $0 < \theta < \pi$ where
\begin{equation}
r_{+} := M + \sqrt{M^2-a^2}  \, \, .
\end{equation} 
As a little reminder on the Kerr geometry we give the following 
basic facts relevant for the discussion of the wave equation. The 
coordinate vector field $\partial / \partial r$ 
becomes singular at 
$r= r_{+}$. This value of the radial coordinate marks the 
event horizon for Kerr spacetime. The coordinate vectorfield
$\partial / \partial t$ is {\it null} on the ergosphere
\begin{equation}
r = M + \sqrt{M^2 - a^2 \cos^2 \theta} \, \, ,
\end{equation} 
is {\it spacelike} inside and {\it timelike} outside.
So $t$ {\it is not a time coordinate inside the ergosphere} and therefore 
one might 
think that Boyer-Lindquist coordinates are unsuitable 
for a stability discussion.   
It turns out that this is not the case for the methods (from semigroup theory)
of this paper. Finally, 
the Kerr metric is globally hyperbolic  outside the horizon
and hence the 
Cauchy problem for the scalar wave equation is well posed for 
data on any Cauchy surface. This result is not used in this paper.
Existence, uniqueness and continuous dependence of the solutions
on the initial data is proved here, too.  
\newline
\linebreak
The reduced wave equation governing solutions
of the form $\psi(t,r,\theta,\phi) = exp(im\varphi)u(t,r,\theta)$, 
where $m$ runs through all integers, is given by
\begin{eqnarray} \label{waveequation}
&&\frac{\partial^2 u}{\partial t^2} + 
\left[\left( \frac{(r^2+a^2)^2}{\triangle} - a^2 \sin^2 \theta \right)^{-1} \cdot 
\right.
\nonumber \\
&& \left. 
\left(i \, \frac{4mMar}{\triangle}\frac{\partial u}{\partial t}  - \frac{\partial}{\partial r} \triangle  \frac{\partial}{\partial r}
- \frac{m^2 a^2}{\triangle} - \frac{1}{\sin \theta} 
\frac{\partial}{\partial \theta} \sin \theta \frac{\partial}{\partial \theta}
+ \frac{m^2}{\sin^2 \theta}\right)\right]u = 0 \, \, .
\end{eqnarray} 
A first inspection shows that 
\begin{equation} \label{basest1}
0 < \frac{4mMar}{\triangle}
\left( \frac{(r^2+a^2)^2}{\triangle} - a^2 \sin^2 \theta \right)^{-1} 
\leqslant \frac{ma}{Mr_{+}}
\end{equation} 
on $\Omega := (r_{+},\infty) \times (0,\pi)$. Hence the coefficient 
multiplying $i \, \partial u / \partial t$ is real-valued, strictly positive and 
bounded. Moreover the coeffcients 
multiplying the derivatives in $r$
and $\theta$ are real-valued and vanish at the horizon. 
As consequence (\ref{waveequation}) {\it 
is singular at all points on the boundary of}
$\Omega$.
\newline
\linebreak
The structure of this paper is as follows.
Section 2 contains the used conventions.
Section 3 gives an initial value formulation  
for (\ref{waveequation}).
There the equation is interpreted as an abstract equation
\begin{equation} \label{waveequation2}
\begin{pmatrix}
u \\ v
\end{pmatrix}^{\prime}
= - \, G 
\begin{pmatrix}
u \\ v 
\end{pmatrix}
= - 
\begin{pmatrix}
-v \\  (A+C)u+iBv 
\end{pmatrix} 
\end{equation} 
for a  differentiable function ${^t}(u,v)$ assuming values in 
an appropriate Hilbert space $Y$ and in particular in the domain of
a linear operator $G$. Here $'$ denote a Hilbert space 
derivative. The linear operators $A+C$ and $B$ will be read off 
from (\ref{waveequation}). $B$ is the maximal 
multiplication operator in $X$ given by the 
function multiplying  $i \, \partial u / \partial t$.
The auxiliar operator
$C$ is a suitable negative multiple of the identity operator on 
$X$. The definition of $A$ is more involved. 
A preliminary form $A_0 + C$ of $A+C$ is given by the 
differential operator enclosed in square brackets.
It has as a domain all complex-valued functions on $\Omega$
which are twice continuously differentiable and have a 
compact support in $\Omega$. Moreover $X$ is chosen such 
that $A_0 + C$ is symmetric. It will be obvious that 
this operator is in addition semibounded (from below). 
In the next step $A+C$ will be defined 
as the {\it Friedrichs extension} of $A_0 + C$. Using this
existence, uniqueness, continuous dependence of
the solutions on the initial data 
(\ref{waveequation2}) follow
from abstract theorems derived in \cite{beyer}.   
In Section 4 the domain of $A$ will be investigated further. 
This is done for two reasons. Firstly, to make 
sure that it contains functions having a reasonable behaviour, both, on  
the axis of symmetry and on the horizon. Secondly, such an
information is needed as a basis for Section 5. There
the resolvent of $G$ 
is constructed 
using spheroidal functions. Section 6 discusses the reduced 
Klein-Gordon equation. The well-posedness
of the initial value problem is shown. Further, the stability 
of the solutions is shown for large enough masses. 
Section 7 contains the discussion. The Appendix gives 
auxiliar theorems used in the computation
of the resolvent.

\section{Conventions}
The symbols ${\Bbb{N}}$, ${\Bbb{R}}$, ${\Bbb{C}}$ denote the 
natural numbers 
(including zero), all real numbers and all
complex numbers, respectively. \newline
\linebreak
To ease understanding we follow common abuse of notation and don't 
differentiate between coordinate maps and coordinates. For instance, 
interchangeably $r$ will denote some real number greater
than $r_{+}$ or the coordinate projection onto the open interval
$(r_{+}, \infty)$. The definition used
will be clear from the context. In addition we assume composition 
of maps (which includes addition, multiplication etc.) always to be   
maximally defined. So for instance the addition of two maps (if at all
possible) is defined on the intersection of the corresponding domains.
\newline
\linebreak
For each $k \in {\Bbb{N}} \setminus \{0\}$, $n \in {\Bbb{N}} \setminus \{0\}$ and 
each non-trivial open subset 
$M$ of ${\Bbb{R}}^n$ the symbol
$C^k(M,{\Bbb{C}})$ denotes the linear space of $k$-times 
continuously differentiable complex-valued functions on $M$. 
Further $C^k_{0}(M,{\Bbb{C}})$ denotes the subspace of
$C^k(M,{\Bbb{C}})$ consisting of those functions which in addition
have a compact support in $\Omega$.
\newline
\linebreak
Throughout the paper Lebesgue integration theory is used
in the formulation of \cite{riesznagy}.  
Compare also Chapter III in 
\cite{hirzebruchscharlau} and Appendix A in \cite{weidmann}.
To improve readability we follow common usage 
and don't differentiate between an 
almost everywhere (with respect to the chosen measure) defined 
function $f$ and the associated equivalence class
(consisting of all almost everywhere defined functions which differ 
from $f$ only on a set of measure zero). 
In this sense
$L_{C}^2\left(M,\rho\right)$, where $\rho$ is some strictly 
positive real-valued continuous function on $M$,     
denotes the Hilbert space of complex-valued, 
square integrable (with respect to the measure $\rho \, d^nx$) 
functions
on $M$. 
The scalar product 
$<| >$ on $L_{C}^2\left(M,\rho \right)$  is defined by
\begin{equation} \label{scalarproduct}
<f|g> := \int_{M}
   f^*
   g \, \rho \,  d^nx  \, \, ,
\end{equation}
for all $f,g \in L_{C}^2\left(M,\rho \right)$, where $^*$ denotes complex 
conjugation on ${\Bbb{C}}$. It is a standard result of functional analysis
that  $C^k_{0}(M,{\Bbb{C}})$ is dense in $L_{C}^2\left(M,\rho\right)$.
\newline
\linebreak
Finally, throughout the paper standard results and nomenclature of 
operator theory is used. For this compare standard 
textbooks on Functional analysis, e.g.,\cite{reedsimon} Vol. I,
\cite{riesznagy,yosida}.

\section{Basic choices and first consequences}
As basic Hilbert space $X$ for  (\ref{waveequation2}) we chose
\begin{equation}
X := L_{C}^2\left(\Omega,g^{00} \sqrt{-|g|}\right) \, \, ,
\end{equation} 
where $|g|$ denotes the determinant of the matrix $g_{ab}$. Note that
\begin{equation}
g^{00} \sqrt{-|g|}  = 
 \left( \frac{(r^2+a^2)^2}{\triangle} - a^2 \sin^2 \theta \right) \sin \theta
\end{equation}
is singular at the horizon. Hence the elements of 
$X$ vanish there in the mean. 
In the limit case $a=0$ this measure reduces to the standard one 
often used for the
stability discussion of the Schwarzschild metric \cite{wald,kay}.
\newline
\linebreak
The operator $B$ is chosen as the maximal multiplication operator
in $X$ by the function multiplying $i \, \partial u / \partial t$
in (\ref{waveequation}). Since that function is bounded  and positive 
real-valued, $B$ is a bounded linear and positive self-adjoint operator
on $X$ given by 
\begin{equation}
Bf =  \frac{4mMar}{\triangle} \left( \frac{(r^2+a^2)^2}{\triangle} - a^2 \sin^2 \theta \right)^{-1} f
\end{equation}
for every $f \in X$. The operator $A_0 + C$ is defined by 
\begin{equation} \label{defA}
(A_0+C)f := D^{2}_{r \theta}f
\end{equation}
for all $f \in C^2_{0}(\Omega,{\Bbb{C}})$, where we set for
every $f \in C^2(\Omega,{\Bbb{C}})$
\begin{eqnarray} \label{D2rtheta}
&& D^2_{r \theta} f :=
\left( \frac{(r^2+a^2)^2}{\triangle} - a^2 \sin^2 \theta \right)
^{-1}\cdot 
\nonumber \\
&& \left( - \frac{\partial}{\partial r} \triangle  \frac{\partial}{\partial r}
- \frac{m^2 a^2}{\triangle} - \frac{1}{\sin \theta} 
\frac{\partial}{\partial \theta} \sin \theta \frac{\partial}{\partial \theta}
+ \frac{m^2}{\sin^2 \theta}\right)f \, \, . 
\end{eqnarray}
Then $A_0 + C$ is in particular 
linear and (using partial integration) symmetric. Further 
again by partial integration
it is easy to see that $A_0 + C$ is semibounded with the lower 
bound $-\alpha$, where $\alpha := m^2 a^2/(4 M^2 r_{+}^2)$. Note that this bound 
approaches $-\infty$ for $|m| \rightarrow \infty$, which would suggest 
that the unreduced wave equation on Kerr background 
would have no stable initial value problem. Also note that it approaches
$0$ for $a \rightarrow 0$ which is the optimal bound for  
Schwarzschild. 
\newline
\linebreak
In the next step we define $C := -(\alpha +\varepsilon)$, where $\varepsilon > 0$
is assumed to have the dimension $l^{-2}$. The exact value of 
$\varepsilon$ does not influence the results in any essential way. 
Finally, we define $A$ as the Friedrichs extension of $A_{0}$.
As a consequence $A$ is a densely-defined, linear, selfadjoint and 
semibounded operator having the same lower bound as $A_{0}$, 
i.e., $\varepsilon$.
\newline
\linebreak
The objects $X,A,B$ and $C$ are easily seen to satisfy Assumptions 1 
and 4 of \cite{beyer}. Applying the results of that paper 
gives 
\begin{theorem}
\begin{description}
\item[(i)] By
\begin{equation} 
Y := D(A^{1/2}) \times X 
\end{equation}
and 
\begin{equation}
(\xi | \eta) := <A^{1/2}\xi_1|A^{1/2}\eta_1> + <\xi_2|\eta_2>
\end{equation}
for all $ \xi = (\xi_1,\xi_2), \eta = (\eta_1,\eta_2) \in Y$
there is defined a complex Hilbert space $(Y,(\, |\, ))$. 
\item[(ii)] The operators
$G$ and $-G$ defined by 
\begin{equation}
G(\xi,\eta) := (-\eta, (A+C)\xi + iB\eta) \, \, 
\end{equation}
for all $\xi \in D(A)$ and $\eta \in D(A^{1/2})$
are infinitesimal generators of strongly continuous semigroups
$T_{+} : [0,\infty) \rightarrow L(Y,Y)$ and 
$T_{-} : [0,\infty) \rightarrow L(Y,Y)$, respectively. 
\item[(iii)]
For all $t \in [0,\infty)$:
\begin{equation} \label{normest}
|T_{\pm}(t)| \leqslant \exp(\|C\| \, \|A^{-1/2}\|t) \, \, \, ,
\end{equation} 
where $|\,|$, $\| \, \|$ denote the operator norm for $L(Y,Y)$ and
$L(X,X)$, respectively.
\item[(iv)]
For every $t_0 \in {\Bbb{R}}$ and every $\xi \in  D(A) \times D(A^{1/2})$ 
there is a uniquely determined 
differentiable map $u : {\Bbb{R}} \rightarrow Y$
such that 
\begin{equation}
u(t_0) = \xi
\end{equation}
and 
\begin{equation} \label{eveq}
u^{\prime}(t) = - G u(t) 
\end{equation}
for all $t \in {\Bbb{R}}$. Here $\, ^{\prime}$ denotes differentiation 
of functions assuming values in $Y$. Moreover this
$u$ is given by 
\begin{equation} \label{rep}
u(t) :=     
 \left\{
 \begin{array}{cl}
 T_{+}(t)\xi & \text{for $t \geqslant 0$} \\
 T_{-}(-t)\xi & \text{for $t < 0$}
 \end{array}
 \right. 
\end{equation}
for all $t \in {\Bbb{R}}$.
\item[(v)] 
$\lambda \in {\mathbb{C}}$ is a spectral value, eigenvalue
of $iG$ if and only if
\begin{equation}
A+C - \lambda B - \lambda^2 
\end{equation}
is not bijective and not injective, respectively.
\item[(vi)] For any $\lambda$ from the resolvent set of $iG$ and any 
$\eta = (\eta_1,\eta_2) \in Y$ one has:
\begin{equation}
(iG - \lambda)^{-1} \eta = \left(\xi,i(\lambda \xi + \eta_1)\right)
\, \, ,
\end{equation}
where 
\begin{equation}
\xi := (A+C - \lambda B - \lambda^2)^{-1} \left[ (B+\lambda)\eta_1 
- i \eta_2\right] \, \, . 
\end{equation}
\end{description}
\end{theorem}
Equation (\ref{eveq}) is the interpretation of (\ref{waveequation})
used in this paper. In this sense (iv) shows the well-posedness of
the initial value problem for (\ref{waveequation}), i.e., the 
existence and uniqueness of the solution and its continuous dependence 
on the initial data. Moreover (\ref{rep}) gives a representation
of the solution and (iii) gives a rough bound for its growth
in time. In general, this bound is not strong enough to imply stability
of the solutions to (\ref{waveequation}). Part (v) reduces the 
determination of the generator's spectrum
to the determination of the spectrum
of the operator polynomial $A + C - \lambda B - \lambda^2 \, , \, \lambda \in 
{\mathbb{C}}$ \cite{markus,rodman}. Moreover (vi) does 
the same for the resolvents. 
Further, \cite{beyer} gives the following stability citeria
\begin{theorem} 
\begin{description}
\item[(i)] If
\begin{equation}
<\xi|(A+C)\xi> + \frac{1}{4} <\xi|B\xi>^2 \, \, \geqslant 0 \nonumber \, \, .
\end{equation}
for all $\xi \in D(A)$ with $\|\xi\|=1$.
Then the spectrum of $iG$ is real.
\item[(ii)] If $A+C - (b/2)B -(b^2/4)$ is positive for some $b \in {\mathbb{R}}$.
Then the spectrum of $iG$ is real and there are $K\geqslant 0$ and 
$t_0 \geqslant 0$ such that 
\begin{equation}
|u(t)| \leqslant Kt  \nonumber
\end{equation}
for all $t \geqslant t_0$. 
\end{description}
\end{theorem}
Here $\| \, \|$, $|\,|$ denote the induced norm on 
$(X, <\,|\,>)$ and $(Y, (\,|\,))$, respectively. Note that 
the reality of the generator's spectrum would exclude 
the existence of exponentially growing mode solutions of (\ref{waveequation}). 
It seems that these criteria are {\it not} strong enough to prove stability
of the solutions of (\ref{waveequation}). 
\footnote{In the following discussion the trivial cases 
$a=0$, i.e., the case of a Schwarschild background, 
and $m=0$ corresponding to purely axial perturbations, are excluded. 
Of course, for these stability of the 
solutions can be concluded from Theorem 2(ii).}
But later on (ii) will be used 
to conclude stability for the corresponding Klein-Gordon equation for cases
where the mass excedes some given bound depending on $m$. Note that  
the positivity of $A_0 + C$ would imply stability via (ii).  
On first sight positivity of  $A_0 + C$ seems unlikely because of the 
negative potential term $- m^2 a^2 / \triangle$ in  
(\ref{D2rtheta}). On the other hand it is well-known that the ocurrence 
of such a term can be due to the chosen representation space 
for $A_0 + C$. In addition the domain of this operator is very much 
restricted by the condition that its elements have a compact support 
in $\Omega$. Since $\Omega$ is open it follows that the support of 
such a function has a strictly positive distance from the boundary. 
In the theory of Schr\"{o}dinger operators it is well-known 
from so called `Poincare-Sobolev inequalities' that the 
kinetic energy associated with such a state can exceed a
negative potential energy. See, e.g., \cite{ziemer} or for a simple 
example \cite{reedsimon} Vol. II, example 1 in chapter X.3.
Indeed such inequalities were found, but only ones leading to 
a positive potential term with asymptotic behaviour $\thicksim
\triangle^{- \beta}$ for $r \rightarrow r_{+}$, where
 $0 \leqslant \beta < 1$. So none of them 
was found to be strong enough to show positivity of $A+C$. Indeed
the apparent absence of better estimates lead to the impression
that $A+C$ is indeed negative. If this is really true it should be
easy to prove using the results  on the domain of $A+C$ from the next 
section. This point has not been 
investigated further, because the negativity alone would not give 
any further information on the stability 
of the solutions of (\ref{waveequation}). 

\section{Investigation of the domain of A+C}

In this Section the domain of $A+C$ will be further investigated. 
This is done for two reasons. Firstly, to make 
sure that it contains functions having a reasonable behaviour, both, 
on the axis of symmetry and on the horizon. It turns out 
that this is indeed the case. In particular, as it should be 
the case, functions of the form $f(r)P^{l}_{m}(\cos \theta)$, where  
$f \in C^{2}_{0}(I_r,{\mathbb{C}})$
and $P^{l}_{m}$, $l=|m|, |m|+1, \dots$, 
are the usual generalized Legendre polynomials 
are found to be in the domain of $A+C$. 
Secondly such an information is needed as a basis for the 
construction of the resolvent of $G$ in the next Section. 
\newline
\linebreak
We do not give a full characterization of $D(A+C)$ here. 
Instead more modestly sufficient conditions are given on 
functions $f(r)$ and $g(\theta)$ which guarantee that the product 
$f(r)g(\theta)$ is in $D(A+C)$. These conditions will turn out to 
be sufficient as a basis for the next Section. They are as follows:
\begin{theorem}
For this denote $I_{r} := (r_{+}, \infty)$ and $I_{\theta} := (0,\pi)$ and 
define 
\begin{equation}
X_r := L_{\Bbb{C}}^2(I_{r},r^4/\triangle) \, \, , \, \,
X_{\theta} :=  L_{\Bbb{C}}^2(I_{\theta},\sin \theta)
\end{equation}
and for every $f \in C^2(I_{r},{\Bbb{C}})$ and
$g \in C^2(I_{\theta},{\Bbb{C}})$ 
\begin{equation}
D^2_{r}f := \frac{\triangle}{r^4} 
\left[-\left(\triangle f^{\, \prime} \right)^{\, \prime} 
- \frac{m^2 a^2}{\triangle} f \right] \, \, , \, \, 
D^2_{\theta} \, g :=
- \frac{1}{\sin \theta} \left( \sin \theta \, g^{\, \prime}\right)^{\, \prime}
+ \frac{m^2}{\sin^2 \theta} \, g \, \, .
\end{equation}
Let be 
$f \in C^2(I_{r},{\Bbb{C}}) \cap X_r$ 
and $g \in  C^2(I_{\theta},{\Bbb{C}}) \cap X_{\theta}$
such that
\begin{equation}
D_{r}^{2}f \in 
X_r \quad \text{and} \quad
D_{\theta}^{2} g \in  
X_{\theta}
\end{equation}
and for $m=0$ in addition such that
\begin{equation} \label{m=0}
\lim_{\theta \rightarrow 0} \, \sin \theta \,  g^{\, \prime}(\theta) = 
\lim_{\theta \rightarrow \pi} \, \sin \theta \, g^{\, \prime}(\theta) = 0 
\, \, .
\end{equation}
Then $f(r)g(\theta) \in D(A+C)$ and 
\begin{equation}
(A+C)f(r)g(\theta) = D^2_{r \theta} f(r)g(\theta) \, \, .
\end{equation}
\end{theorem}
{\bfseries Proof:}
First it follows from the obvious inequalities
\begin{equation} \label{basest2}
\frac{r^4}{\triangle} \leqslant
\frac{(r^2+a^2)^2}{\triangle} - a^2 \sin^2 \theta
\leqslant \frac{4M^2}{r_{+}^2} \,
\frac{r^4}{\triangle}  \, \, , \, \,
\end{equation}
that   
$L_{\Bbb{C}}^2(\Omega, r^4 \sin \theta / \triangle)$ and $X$
are identical as sets and that the associated norms on that set
are equivalent. A further consequence of (\ref{basest2}) along with
partial integration is the fact that $f(r)g(\theta)$ is in the domain
$D((A_0+C)^{*})$ of the adjoint $(A_0 + C)^{*}$ of $A_0 + C$
and in particular that
\begin{equation}
(A_0 + C)^{*}f(r)g(\theta) = D^2_{r \theta} f(r)g(\theta) \, \, .
\end{equation}
Hence $f(r)g(\theta) \in D((A_0+C)^{*})$ if and only if there is a 
sequence $h_0, h_1, \cdots$ of elements of $C^2_{0}(\Omega,{\Bbb{C}})$ 
converging to 
$f(r)g(\theta)$ and such that for every given $\varepsilon > 0$ there is  
$\nu_0 \in {\Bbb{N}}$ such that for all $\mu, \nu \in {\Bbb{N}}$
satisfying $\mu \geqslant \nu_0$ and  $\nu \geqslant \nu_0$:
\begin{equation} \label{Cauchy}
|<h_{\mu}- h_{\nu}|(A_{0}+C + \alpha)(h_{\mu}- h_{\nu})>| \, < \, \varepsilon \, 
\, .
\end{equation}
In the following the existence of such a sequence will be shown. Basic for 
this is the following inequality valid for all 
$u \in C_{0}^2(I_{r},{\Bbb{C}})$ and $v \in C_{0}^2(I_{\theta},{\Bbb{C}})$:
\begin{eqnarray} 
&&<u(r)v(\theta)|(A_0+C+\alpha)u(r)v(\theta)> \, \, \leqslant
\left( \int_{r_{+}}^{\infty} |u|^2 dr \right)
\left( \int_{0}^{\pi} \sin \theta  \,
v^{*} D^2_{\theta} v \, d\theta \right) +  \nonumber \\
&&\left( \int_{r_{+}}^{\infty} r^4 \triangle^{-1}
u^{*} (D^2_{r}+m^2 a^2/r_{+}^4) u \, dr \right)
\left( \int_{0}^{\pi} \sin \theta \, |v(\theta)|^2 
  \, d\theta \right) \leqslant \\
&&\left( \int_{r_{+}}^{\infty} r^4 \triangle^{-1}
u^{*} (D^2_{r} + m^2 a^2/r_{+}^4+ r_{+}^{-2}) u \,  dr \right)
\left( \int_{0}^{\pi} \sin \theta  \,
v^{*} (D^2_{\theta}+1) v \, d\theta \right) \, \, . \nonumber
\end{eqnarray} 
Here some elementary estimates have been used along 
with the positivity of $D^2_{r} + m^2 a^2/r_{+}^4$ 
on $ C_{0}^2(I_{r},{\Bbb{C}}) \subset X_r$.
Since $A+C+\alpha$ is in particular positive also the following
inequality is valid for all $u_1, u_2 \in C_{0}^2(I_{r},{\Bbb{C}})$ 
and $v_1,v_2 \in C_{0}^2(I_{\theta},{\Bbb{C}})$:
\begin{eqnarray} \label{factor}
&&|<u_1(r)v_1(\theta)-u_2(r)v_2(\theta)|(A_0+C+\alpha)
[u_1(r)v_1(\theta)-u_2(r)v_2(\theta)]>| \nonumber \\ 
&& = \|(A+C+\alpha)^{1/2}[u_1(r)-u_2(r)]v_1(\theta)+
        (A+C+\alpha)^{1/2}u_2(r)[v_1(\theta)-v_2(\theta)]\|^2 \nonumber \\
&& \leqslant 
2 <[u_1(r)-u_2(r)]v_1(\theta)|
(A_0+C+\alpha)[u_1(r)-u_2(r)]v_1(\theta)> +  \nonumber \\
&& \quad 2 <u_2(r)[v_1(\theta)-v_2(\theta)]|(A_0+C+\alpha)u_2(r)[v_1(\theta)-v_2(\theta)]>
\end{eqnarray}
where $\alpha^{\, \prime} :=  m^2 a^2/r_{+}^4+ r_{+}^{-2}$. 
Since $f$ is in the domain of the Friedrichs extension 
of $D^2_{r}$ on $ C_{0}^2(I_{r},{\Bbb{C}}) \subset X_r$ 
there is a sequence 
$f_0, f_1, \cdots$ of elements of $C^2_{0}(I_{r},{\Bbb{C}})$ 
converging to 
$f(r)$ and such that for every given $\varepsilon > 0$ there is  
$\nu_0 \in {\Bbb{N}}$ such that for all $\mu, \nu \in {\Bbb{N}}$
satisfying $\mu \geqslant \nu_0$ and  $\nu \geqslant \nu_0$:
\begin{equation} \label{Cauchyr}
\int_{r_{+}}^{\infty} r^4 \triangle^{-1}
\left(f_\mu-f_\nu\right)^{*} 
(D^2_{r} + \alpha^{\, \prime}) (f_\mu - f_\nu)\, 
 dr \, < \, \varepsilon \, 
\, .
\end{equation}
Obviously, by an argument analogous to (\ref{factor}) this implies that
the sequence 
\begin{equation}
\int_{r_{+}}^{\infty} r^4 \triangle^{-1}
f_\nu^{*} 
(D^2_{r} + \alpha^{\, \prime}) f_\nu \, 
 dr \, \, , \, \, \nu \in {\Bbb{N}}
\end{equation}  
is bounded. 
Moreover since $g$ is in the domain of the Friedrichs extension 
of $D^2_{\theta}$ on $ C_{0}^2(I_{\theta},{\Bbb{C}}) \subset X_\theta$ 
there is a sequence 
 $g_0, g_1, \cdots$ of elements of $C^2_{0}(I_{\theta},{\Bbb{C}})$ 
converging to 
$g(\theta)$ and such that for every given $\varepsilon > 0$ there is  
$\nu_0 \in {\Bbb{N}}$ such that for all $\mu, \nu \in {\Bbb{N}}$
satisfying $\mu \geqslant \nu_0$ and  $\nu \geqslant \nu_0$:
\begin{equation} \label{Cauchytheta}
\int_{0}^{\pi} \sin \theta \,
\left(g_\mu-g_\nu\right)^{*} 
(D^2_{\theta} + 1) \, (g_\mu - g_\nu) \,
d \theta \, < \, \varepsilon \, 
\, .
\end{equation}
Here too, by an argument analogous to (\ref{factor}) this implies that
the sequence 
\begin{equation}
\int_{0}^{\pi} \sin \theta \,
g_\nu^{*} \, 
(D^2_{\theta} + 1) \, g_\nu \, 
d\theta \, \, , \, \, \nu \in {\Bbb{N}}
\end{equation}  
is bounded. 
Finally, because of
\begin{eqnarray}
&&|<f_\mu(r)g_\mu(\theta)-f_\nu(r)g_\nu(\theta)|(A_0+C+\alpha)
[f_\mu(r)g_\mu(\theta)-f_\nu(r)g_\nu(\theta)]>| \leqslant \nonumber \\ 
&& 2 \left( \int_{r_{+}}^{\infty} r^4 \triangle^{-1}
\left(f_\mu-f_\nu\right)^{*} 
(D^2_{r} + \alpha^{\, \prime}) (f_\mu - f_\nu)\, dr \right)
\left( \int_{0}^{\pi} \sin \theta  \,
g_\mu^{*} (D^2_{\theta}+1) g_\mu \, d\theta \right) + \nonumber \\
&& 2 \left( \int_{r_{+}}^{\infty}  r^4 \triangle^{-1}
f_\nu^{*} (D^2_{r} + \alpha^{\, \prime}) f_\nu \, dr 
\right) \left( \int_{0}^{\pi} \sin \theta  \,
\left(g_\mu-g_\nu \right)^{*} (D^2_{\theta}+1) \left(g_\mu-g_\nu
\right)\, d\theta \right) \, \, ,
\end{eqnarray}
the sequence $h_0, h_1, \dots$ defined by 
\begin{equation}
h_{\nu} := f_{\nu}(r)g_{\nu}(\theta) \, \, , \, \, \nu \in {\Bbb{N}}
\end{equation}
has the required properties.$_{\square}$
\newline
\linebreak
In the proof there have been used facts on the Sturm-Liouville
operators $D_{r}^{2}$ and $D_{\theta}^{2}$. Now, for the reader's 
convenience these will be given. 
For this define the (obviously) 
linear, symmetric and semibounded operators $A_{r0}$, 
$A_{\theta 0}$ in $X_r$ and $X_{\theta}$, respectively, by
\begin{equation}
A_{r0}f := D_r^{2}f \, \, , \, \, 
A_{\theta 0}g := D_{\theta}^{2}g \, \, , 
\end{equation}
for every $f \in C_{0}^2(I_{r},{\Bbb{C}})$ and every
$g \in C_{0}^2(I_{\theta},{\Bbb{C}})$.
Then one has the following 
\begin{lemma}
\begin{description}
\item[(i)]  $A_{r0}$ is essentially self-adjoint.
\item[(ii)] $A_{\theta 0}$ is essentially self-adjoint
for $m>0$. For $m=0$, the Friedrichs extension
of $A_{\theta 0}$ is given by  
the closure of the operator $A_{\theta F}$ defined
by $A_{\theta F}f := D_{\theta}^{2}g$ for every 
$g \in  C^2(I_{\theta},{\Bbb{C}}) \cap X_{\theta}$
satisfying (\ref{m=0}) together with the condition that  
$D_{\theta}^{2} g \in  X_{\theta}$. For all $m$ the spectrum 
of the Friedrichs extension of  $A_{\theta 0}$ is given 
by $\{|m|(|m|+1),(|m|+1)(|m|+2), \dots \}$.  
\end{description}
\end{lemma}
{\bfseries Proof:}
(i) For this define the auxiliar Sturm-Liouville operator 
$\hat{A}_{r \, 0}$ in $X_{r}$ by 
\begin{equation}
\hat{A}_{r \, 0}f := -\frac{\triangle}{r^4} 
\left(\triangle f^{\, \prime} \right)^{\, \prime} 
\end{equation}
for every $f \in C_{0}^2(I_{r},{\Bbb{C}})$. Obviously,
$\hat{A}_{r \, 0}$ is densely-defined, linear,
symmetric and positive. Moreover since $-m^2a^2/r^4$ is bounded
on $I_{r}$, it follows by the
Rellich-Kato theorem (see, e.g, Theorem X.12 in Volume II of
\cite{reedsimon}) that  $A_{r0}$ is essentially self-adjoint
if and only if $\hat{A}_{r0}$ is essentially self-adjoint.
Now, the equation $\left(\triangle f^{\, \prime} \right)^{\, \prime}=0$
has nonvanishing constants as solutions. Since these are not in $X_{r}$
at both ends of $I_r$, it follows that $\hat{A}_{r0}$ is in the 
limit point case, both, at $r_{+}$ and at $+\infty$. Hence $\hat{A}_{r0}$
is essentially self-adjoint (see, e.g., \cite{weidmann}). Finally, from this
follows (i). 
(ii) This statement is, of course, well-known. $_{\square}$

\section{Computation of the generator's resolvent}

In the following the resolvent of $G$ will be determined 
for spectral parameters $\lambda$ which are 
non real and at the same time such that $ia\lambda$
is not an exceptional value.\footnote{For the definition of these values see
\cite{meixner}.} Note that, because of Theorem 1 (vi), the resolvent 
of $G$ can be derived from the inverses of the operator polynomial
$A+C -\lambda B -\lambda^2$ which are given in (ii) of the following theorem
on a dense subset of $X$. 
\begin{theorem}
Let $\lambda$ be a non real element of the resolvent set of $iG$ which
moreover
is such that $ia\lambda$ is not an exceptional value. For each
$m \in {\mathbb{Z}}$ let
\begin{equation}
ps^{m}_{l}(\cos \theta, -a^2 \lambda^2) \, \, 
, \, \, l = |m|,|m|+1,|m|+2, \cdots 
\end{equation}
be the basis 
\footnote{In the sense that 
the span of these
functions is dense in $X_{\theta}$. Note that these functions are 
{\it not} orthogonal in general. Instead this sequence 
and the sequence consisting of its complex 
conjugates form a {\it biorthogonal} Basis of $X_{\theta}$.
See Theorem 4 in Section 3.23 of \cite{meixner}. } 
of $X_{\theta}$ consisting of spheroidal eigenfunctions
of $D^2_{\theta} + \lambda^2 a^2 \sin^2 \theta$ corresponding to 
the eigenvalues
\begin{equation}
\lambda^{m}_{|m|}(-a^2 \lambda^2),\lambda^{m}_{|m|+1}(-a^2 \lambda^2)
,\dots \, \, ,
\end{equation} 
respectively.\footnote{For 
the definition of the functions $ps_{l}^{m}$ see \cite{meixner}.} 
Finally, let
be $g \in C_0(I_r,{\mathbb{C}})$, $m \in {\mathbb{Z}}$
and $ l \in \{|m|,|m|+1,\dots\}$. Then
\begin{description}
\item[(i)] The subset of $X$ consisting of all finite linear combinations
of elements of the form  
\begin{equation}
h(r,\theta)g(r)p^{m}_{l}(\cos \theta, -a^2 \lambda^2)
\end{equation}
where 
\begin{equation}
h := \frac{r^4}{\triangle}
\left( \frac{(r^2+a^2)^2}{\triangle} - a^2 \sin^2 \theta \right)^{-1} \, \, ,
\end{equation}
and $g$, $l$ run through the elments of $C_0(I_r,{\mathbb{C}})$ and 
$\{|m|,|m|+1,|m|+2, \dots\}$, respectively,
is dense in $X$.
\item[(ii)]
\begin{eqnarray} \label{inverse}
&&(A+C - \lambda B - \lambda^2)^{-1}
h(r,\theta)g(r)p^{m}_{l}(\cos \theta, -a^2 \lambda^2) = \nonumber \\
&&f_{r}(r)p^{m}_{l}(\cos \theta, -a^2 \lambda^2) \, \, , 
\end{eqnarray}
where $f_{r} \in C^2(I_{r},{\Bbb{C}}) \cap X_r$ is such  
that $D_{r}^{2}f_{r} \in X_{r}$ and moreover satisfies
\begin{equation} \label{inversion}
D^2_{r\lambda}f_{r}+ 
\lambda^{m}_{l}(-a^2 \lambda^2)(\triangle/r^4) f_{r} = g \, \, .
\end{equation}
Here for every 
$\phi \in C^2(I_{r},{\Bbb{C}})$
\begin{equation}
D^2_{r\lambda} \phi :=
-\frac{\triangle}{r^4}\left( \triangle {\phi}^{\, \prime} \right)^{\, \prime} -
\frac{1}{r^4}\left[( ma + 2\lambda M r)^2 +  
 \lambda^2 \triangle (\triangle + 4Mr)\right]{\phi} \, \, . 
\end{equation}
\end{description}
\end{theorem}
{\bfseries Proof:} First we notice that $h$ 
is $C^{\infty}$ on $\Omega$ and satisfies
as a consequence of (\ref{basest2})
\begin{equation}
r_{+}^2/(4M^2) \leqslant h \leqslant 1 \, \, .
\end{equation}  
Hence the maximal multiplication operator $T_{h}$ by $h$ in $X$
is defined on the whole of $X$, is bijective and its inverse
is given by the maximal multiplication operator $T_{1/h}$ 
which is defined on the whole of $X$, too. 
by the function $1/h$ in $X$. 
Obviously, the subset of $X$ consisting of all finite linear combinations
of elements of the form  
\begin{equation}
g(r)p^{m}_{l}(\cos \theta, -a^2 \lambda^2)
\end{equation}
where $g \in  C_0(I_r,{\mathbb{C}})$ and $l = |m|,|m|+1,|m|+2, \dots$
is dense in $X$. Hence this is true for 
the subset of $X$ consisting of all finite linear combinations
of elements of the form
\begin{equation}   
h(r,\theta)g(r)p^{m}_{l}(\cos \theta, -a^2 \lambda^2)
\end{equation}
where $g \in  C_0(I_r,{\mathbb{C}})$ and $l = |m|,|m|+1,|m|+2, \dots$, too.
In the following let  $g$ be some element of $C_0(I_r,{\mathbb{C}})$ 
and $l$ be some element of $\{|m|,|m|+1,|m|+2, \dots\}$. 
We will compute the element $f \in X$ satisfying 
\begin{equation}
\left(A+C - \lambda B -\lambda^2 \right)f(r,\theta) = 
h(r,\theta)g(r)p^{m}_{l}(\cos \theta, -a^2 \lambda^2) \, \, .
\end{equation} 
We note that by Theorem 2 
\begin{eqnarray}
&&\left(A+C - \lambda B -\lambda^2 \right)f_{r}(r)
p^{m}_{l}(\cos \theta, -a^2 \lambda^2) = \nonumber \\
&&h(r,\theta) 
\left[ \left(D^2_{r\lambda}f_{r}+ 
\lambda^{m}_{l}(-a^2 \lambda^2)(\triangle/r^4) f_{r}
\right)(r)  
\right] 
p^{m}_{l}(\cos \theta, -a^2 \lambda^2)
\end{eqnarray}
for every 
$f_{r} \in C^2(I_{r},{\Bbb{C}}) \cap X_r$ 
such that
\begin{equation}
D_{r}^{2}f_{r} \in X_{r} \, \, . 
\end{equation}
In the following we construct such a $f_{r}$ which satisfies in particular
(\ref{inversion})
Then by the bijectivity of $A+C - \lambda B -\lambda^2$ we conclude that
\begin{equation}
f(r,\theta) = f_{r}(r)
p^{m}_{l}(\cos \theta, -a^2 \lambda^2) \, \, .
\end{equation} 
For this construction we need some auxiliar solutions
$f_1$, $f_2$, $f_3$ and $f_4$ of the homogeneous
equation associated with (\ref{inversion}), i.e.,
\begin{equation} \label{inversion1}
f_{r}^{\, \prime \prime} + 
\frac{2(r-M)}{\triangle} f_{r}^{\, \prime}+
\left[\frac{(ma +2\lambda Mr)^2}{\triangle^2}
+ \lambda^2
\left(1+\frac{4Mr}{\triangle}\right)+\frac{s}{\triangle}\right]f_{r} = 0 \, \, ,
\end{equation}
where $s := \lambda^{m}_{l}(-a^2 \lambda^2)$,
having special asymptotic behaviour at the singular point $r=r_{+}$
and at $+\infty$.
First, by defining $\bar{f}_{r} := \triangle^{1/2}f_{r}$ and by introducing 
the new
independent variable $r_{*}$ 
\begin{equation}
r_{*} := \sqrt{r(r+4M)} + 2M\ln\left((\sqrt{r+4M}+\sqrt{r})^2/(4M)\right)
\end{equation}
one gets a homegeneous first order system for 
$\bar{f}_{r}$ and $d\bar{f}_{r}/dr_{*}$ which is equivalent to 
(\ref{inversion1}) and which satisfies the assumptions of Theorem 4 in the 
appendix. From this theorem follows the existence of 
linear independent continuously differentiable 
solutions $(\bar{f}_{r1},d \bar{f}_{r1}/dr_{*})$ 
and $(\bar{f}_{r2},d \bar{f}_{r2}/dr_{*})$ 
of the system
along with continuously differentiable functions $R_{1}$ and 
$R_{2}$
such that
\begin{eqnarray} \label{asymptotic1}
&& \bar{f}_{r1}(r_{*}) = e^{i \lambda r_{*}} \left( 1 + R_{11}(r_{*})\right)
\, \, , \, \, 
\frac{d\bar{f}_{r1}}{d r_{*}}(r_{*}) 
= e^{i \lambda r_{*}} \left( i \lambda + R_{12}(r_{*})\right) \nonumber \\
&& \bar{f}_{r2}(r_{*}) = e^{-i \lambda r_{*}} \left( 1 + R_{21}(r_{*})\right)
\, \, , \, \, 
\frac{d\bar{f}_{r2}}{d r_{*}}(r_{*}) 
= e^{-i \lambda r_{*}} \left( - i \lambda + R_{22}(r_{*})\right) \nonumber \\
&& \lim_{r_{*} \rightarrow \infty} |R_{1}(r_{*})| =  
\lim_{r_{*} \rightarrow \infty} |R_{2}(r_{*})| = 0 \, \, .
\end{eqnarray}
In the following denote by $f_{r1}$, $f_{r2}$ the solutions of
(\ref{inversion1}) corresponding to  $(\bar{f}_{r1},d \bar{f}_{r1}/dr_{*})$ 
and $(\bar{f}_{r2},d \bar{f}_{r2}/dr_{*})$, respectively.
Morover define 
\begin{equation}
f_{rR} := 
\left\{
 \begin{array}{cl}
 f_{r1} & \text{for $Im(\lambda) > 0$} \\
 f_{r2} & \text{for $Im(\lambda) < 0$ \, .}
 \end{array}
 \right.
\end{equation} 
Then it follows by (\ref{asymptotic1}) that 
$\phi f_{rR} \in C^2(I_{r},{\Bbb{C}}) \cap X_r$ 
and $D_{r}^{2}(\phi f_{rR}) \in X_r$ for every 
$\phi \in C^2(I_{r},{\Bbb{R}})$ which is identically $0$ for 
$r < r_{0}$ and identically $1$ for $r > r_{1}$, where 
$ r_{0}, r_{1} \in I_{r}$ are such that $r_{0} < r_{1}$, but 
otherwise arbitrary.
For the second step by defining $g_{1} := f_{r} / \triangle$
and $g_{2} := f^{\, \prime}$ one gets a homegeneous first order system for 
$(g_{1},g_{2})$ which is equivalent to 
(\ref{inversion1}) and which satisfies the assumptions of Corollary 5 in the 
appendix. From this corollary follows the existence of 
linear independent continuously differentiable 
solutions $(g_{11},g_{12})$ 
and  $(g_{21},g_{22})$  
of the system
along with continuously differentiable functions $R_{3}$ and 
$R_{4}$
such that
\begin{eqnarray} \label{asymptotic2}
&& g_{11}(r) = (r-r_{+})^{- \sigma_{1}} [1 + R_{31}(r)] \, \, , \, \,
g_{12}(r) = (r-r_{+})^{- \sigma_{1}} [-i (ma + 2\lambda M r_{+}) + R_{32}(r)] 
\nonumber \\
&& g_{21}(r) = (r-r_{+})^{- \sigma_{2}} [1 + R_{41}(r)] \, \, , \, \,
g_{22}(r) = (r-r_{+})^{- \sigma_{2}} [i (ma + 2\lambda M r_{+}) + R_{42}(r)] 
\nonumber \\
&& \lim_{r_{*} \rightarrow r_{+}} |R_{3}(r_{*})| =  
\lim_{r_{*} \rightarrow r_{+}} |R_{3}(r_{*})| = 0 \, \, ,
\end{eqnarray}
where
\begin{eqnarray}
\sigma_1 &:=& 
\left[ \sqrt{M^2-a^2} + i\left((ma/2)+ \lambda M r_{+}
\right)
\right] / \sqrt{M^2-a^2} \nonumber \\
\sigma_2 &:=& 
\left[ \sqrt{M^2-a^2} - i\left((ma/2)+ \lambda M r_{+}
\right)
\right] / \sqrt{M^2-a^2} \, \, .
\end{eqnarray}
In the following denote by $f_{r3}$, $f_{r4}$ the solutions of
(\ref{inversion1}) corresponding to $(g_{11},g_{12})$ 
and $(g_{21},g_{22})$, respectively. 
Moreover define 
\begin{equation}
f_{rL} := 
\left\{
 \begin{array}{cl}
 f_{r3} & \text{for $Im(\lambda) > 0$} \\
 f_{r4} & \text{for $Im(\lambda) < 0$ \, .}
 \end{array}
 \right.
\end{equation} 
Then it follows by (\ref{asymptotic2}) that 
$\phi f_{rL} \in C^2(I_{r},{\Bbb{C}}) \cap X_r$ 
and $D_{r}^{2}(\phi f_{rL}) \in X_r$ for every 
$\phi \in C^2(I_{r},{\Bbb{R}})$ which is identically $1$ for 
$r < r_{0}$ and identically $0$ for $r > r_{1}$, where 
$ r_{0}, r_{1} \in I_{r}$ are such that $r_{0} < r_{1}$, but 
otherwise arbitrary.
\newline
\linebreak
In the next step we notice that $f_{rR}$ and  $f_{rL}$ are linear 
independent, because otherwise we would get a contradiction 
to the assumed bijectivity of $A+C - \lambda B -\lambda^2$. 
Hence the Wronski determinant $W$ of $f_{rR}$ and  $f_{rL}$
\begin{equation}
W := \triangle \left( 
f_{rL}\, f_{rR}^{\, \prime} - f_{rL}^{\, \prime} \, f_{rR} \right)
\end{equation} 
is constant and different from $0$. Therefore we can define
\begin{equation}
f_{r}(r) := 
-  \frac{f_{rR}(r)}{W} \int_{r_{+}}^{r} f_{rL}(r^{\, \prime}) 
g(r^{\, \prime}) dr^{\, \prime}
-  \frac{f_{rL}(r)}{W} \int_{r}^{\infty} f_{rR}(r^{\, \prime}) 
g(r^{\, \prime}) dr^{\, \prime} 
\end{equation}
for all $r \in I_{r}$. 
\newline
\linebreak
Using the foregoing results on 
$f_{rL}$ and $f_{rR}$ by a simple computation
It follows from the foregoing results on 
$f_{rL}$ and $f_{rR}$ and from a simple computation 
that $f_{r} \in C^2(I_{r},{\Bbb{C}}) \cap X_r$,
$D_{r}^{2}f_{r} \in X_{r}$ and that $f_{r}$ satiesfies (\ref{inversion}).
Finally, from the bijectivity of  $A+C - \lambda B -\lambda^2$
we conclude (\ref{inverse}).$_{\square}$

\section{The case of the Klein-Gordon equation}

Compared to the wave equation considered in the previous Sections,
the only change in this case is that the operator 
$C$ has to be substituted by $C^{\, \prime} := C + (m_0^2/g^{00})$, where $m_0$ 
denotes the mass of the field and $m_0^2/g^{00}$ is the maximal 
multiplication operator in $X$, which 
is defined on the whole of $X$ as well as bounded, 
since this function is easily seen to be bounded on $\Omega$. The other
objects $X$, $A$ and $B$ stay the same. Again it is easy to verify that 
$X$, $A$, $B$ and $C^{\, \prime}$ satisfy Assumptions 1 and
4 of \cite{beyer}. As a consequence one has theorems analogous to Theorem 1
and Theorem 2. They imply the well-posedness of the initial value
problem, i.e., the existence and uniqueness of the solution and its continuous
dependence on the initial data. Further, via the analogue of Theorem 2 (ii),
Theorem 7 below implies for masses satisfying  
(\ref{massineq}), that the spectrum of the corresponding generator is 
real and that the norm of the solutions grow at most linearly in time. 
In particular there are no exponentially growing modes in these cases.
\begin{lemma}
Let $B^{\, \prime}$ be a bounded linear and self-adjoint operator on X. Then
$A+B^{\, \prime}$ is identical to the Friedrichs extension
of $A_0+B^{\, \prime}$.    
\end{lemma}
{\bfseries Proof:}
First, since  $B^{\, \prime}$ is bounded linear and self-adjoint 
on X, it follows that
\begin{equation} \label{adjoints}
(A_0 + B^{\, \prime})^{*} = A_0^{*} + B^{\, \prime} \, \, .
\end{equation}
Hence the domain of the Friedrichs extension $(A_0+B^{\, \prime})_F$ of 
$A_0+B^{\, \prime}$ is given by those elements $f$ from $D( A_0^{*})$
for which there is a sequence $f_0,f_1,\dots$ in $D(A_0)$
converging to $f$ and such that for every $\delta > 0$ there is a corresponding 
$\nu_0 \in {\mathbb{N}}$ such that for all $\mu,\nu \in  {\mathbb{N}}$
\begin{equation} \label{fineq}
|<f_{\mu}-f_{\nu}|(A_0 + B^{\, \prime } + 
\|B^{\, \prime}\|)(f_{\mu}-f_{\nu})>| \, \, < \, \, \delta 
\end{equation}   
if, both, $\mu > \mu_0$ and $\nu > \nu_0$.
Since (\ref{fineq}) implies  
\begin{equation}
|<f_{\mu}-f_{\nu}|A_0(f_{\mu}-f_{\nu})>| \, \, < \, \, \delta
\end{equation}
it follows that $f$ is an element of $D(A)$, too. Further,
(\ref{adjoints}) implies 
\begin{equation}
(A_0+B^{\, \prime})_{F}f = (A_0 + B^{\, \prime})^{*}f = Af + B^{\, \prime}f 
\, \, .
\end{equation} 
Hence $A+ B^{\, \prime}$ is a linear self-adjoint (by the
Rellich-Kato theorem, see, e.g, Theorem X.12 in Volume II of
\cite{reedsimon}) extension of
$(A_0+B^{\, \prime})_F$ and, 
finally, since  $(A_0+B^{\, \prime})_F$ is self-adjoint,
$(A_0+B^{\, \prime})_F = A+ B^{\, \prime}$.${_\square}$
\begin{theorem} Define $b := ma/(Mr_{+})$ and 
let be
\begin{equation} \label{massineq}
m_0 \geqslant \frac{|m|a}{2Mr_{+}} \sqrt{1 + \frac{2M}{r_{+}}+ \frac{a^2}{r_{+}^2}}
\, \, .
\end{equation}
Then
\begin{equation}
A + C  + m_0^2/g^{00} + (b/2)B - b^2/4 
\end{equation} 
is positive.
\end{theorem}
{\bfseries Proof:} Because of the preceeding Lemma it is enough
to prove that
\begin{equation}
<f|(A_0 + C + m_0^2/g^{00} + (b/2)B - b^2/4)f> \, \, \geqslant \, \, 0
\end{equation}
for all  $f \in C^2_{0}(\Omega,{\Bbb{C}})$.
Now let $f$ be such an element. Since
its support $supp(f)$ 
is a compact subset of $\Omega$ there are $r_0 > r_{+}$
and $r_1 > r_0$ such that $supp(f) \subset J \times (0,\pi) \subset
\Omega$, where $J := (r_0,r_1)$.
In a first step one gets by partial integration, 
Fubini's theorem and Lemma 4 (ii) 
\begin{eqnarray} \label{est}
&& <f|(A_0+C)f> \nonumber \\ 
&& = \int_{\Omega} \sin \theta \, f^{*}
\left( - \frac{\partial}{\partial r} \triangle  \frac{\partial}{\partial r}
- \frac{m^2 a^2}{\triangle} - \frac{1}{\sin \theta} 
\frac{\partial}{\partial \theta} \sin \theta \frac{\partial}{\partial \theta}
+ \frac{m^2}{\sin^2 \theta}\right)f
 \, dr \, d\theta \, \, \nonumber  \\
&&= \int_{0}^{\pi} \left[ \int_{r_0}^{r_1} 
f_{\theta}^{*} \left( -\frac{d}{dr}\triangle\frac{d}{dr}
-\frac{m^2 a^2}{\triangle}\right)f_{\theta} \, dr \right] \sin \theta \, d \theta 
\nonumber \\
&& \quad + \int_{r_0}^{r_1} \left[
\int_{0}^{\pi} \sin \theta \, f_{r}^{*}
\left(- \frac{1}{\sin \theta} 
\frac{d}{d \theta} \sin \theta \frac{d}{d \theta}
+ \frac{m^2}{\sin^2 \theta}\right)f_{r} \, d \theta
\right] dr \nonumber \\
&&\geqslant
\int_{0}^{\pi} \left[ \int_{r_0}^{r_1}
f_{\theta}^{*} \left( |m|(|m|+1)  
-\frac{m^2 a^2}{\triangle}\right)f_{\theta} \, dr \right] 
\sin \theta \, d \theta \, \, . \nonumber 
\end{eqnarray} 
Further using 
\begin{eqnarray}
<f|f/g^{00}> \quad &\geqslant& \quad  \int_{0}^{\pi} \left[
\int_{r_0}^{r_1} r^2 \, 
|f_{\theta}|^2 \, dr \right] \sin \theta \, \, d\theta \, \, , \\
<f|Bf> \quad &=& \quad 4mMa \int_{0}^{\pi} \left[
\int_{r_0}^{r_1} \frac{r}{\triangle} \, 
|f_{\theta}|^2 \, dr \right] \sin \theta \, \, d\theta \\ 
<f|f> \quad &\leqslant& \quad \int_{\Omega}
\frac{(r^2+a^2)^2}{\triangle} \,  
\sin \theta \, |f|^2 \, dr \, d\theta
\end{eqnarray}
we get
\begin{eqnarray}
&& <f|\left(A_0 + C  + m_0^2/g^{00} + (b/2)B - b^2/4\right)f>  \nonumber\\
&& \geqslant  \int_{0}^{\pi} \left[ \int_{r_0}^{r_1}
f_{\theta}^{*} \left( |m|(|m|+1)  
-\frac{m^2 a^2}{r_{+}^2} \frac{r-r_{+}}{r-r_{-}}
+ m_{0}^2 r^2 \right.  \right. \nonumber \\
&& \quad \quad \quad \quad \quad \quad \quad \, 
\left. \left.  - \frac{m^2 a^2}{4 M^2 r_{+}^2} (r^2+2Mr+a^2) \right)
f_{\theta} \, dr \right] 
\sin \theta \, d \theta  \nonumber \\
&& \geqslant
\int_{0}^{\pi} \left[ \int_{r_0}^{r_1}
f_{\theta}^{*} \left( |m| + m^2 \left( 1 - \frac{a^2}{r_{+}^2} \right)
\right. \right. \\
&& \quad \left. \left. 
+ \frac{m^2 a^2}{4M^2 r_{+}} \cdot \left[
(r_{+}^2 + 2Mr_{+}+a^2)r^2 - r_{+}^2 (r^2+2Mr+a^2)
\right]   
\right)f_{\theta} \, dr \right] 
\sin \theta \, d \theta  \geqslant 0 \, \, . \nonumber 
\end{eqnarray}
 Hence it 
follows the positivity of $A_0 + C  + m_0^2/g^{00} + (b/2)B - b^2/4$.$_\square$
\section{Discussion}
The reduced (in the angular coordinate $\varphi$) 
wave equation and Klein-Gordon equation 
were considered on a Kerr background and in the 
framework of $C^{0}$-semigroup theory. Each equation
was shown to have a well-posed initial value problem,
i.e., to have a unique solution depending 
continuously on the data. Further, it was proved that the 
spectrum of the semigroup's generator coincides with the 
spectrum of an operator polynomial whose coefficients can 
be read off from the equation. In this way the problem of 
deciding stability is reduced to a spectral problem. In addition
a mathematical basis is provided for mode considerations.
\footnote{To give an example for this claim, say, we would be 
able to show that the unstable spectrum of $G$ consists of 
discrete eigenvalues and that the corresponding eigenstates
seperate in the way assumed by Whiting. Then, via the results 
of this paper, Whiting's result \cite{whiting} on the absence of 
exponentially growing modes would imply the stability of the 
solutions of the wave equation.}  
For the wave equation it was shown that the resolvent of 
the semigroup's generator and the corresponding Green's 
functions can be computed using spheroidal functions. 
It is to be expected that,
analogous to the case of a Schwarzschild background,  
the quasinormal frequencies of the Kerr black hole 
appear as {\it resonances}, i.e., poles of the analytic continuation of this resolvent. 
Finally, stability of the background with respect to 
reduced massive perturbations was proved for 
masses exceeding a given bound (see (\ref{massineq})).
\newline
\linebreak
It is interesting to compare the last result to 
earlier results of Detweiler in \cite{detweiler}, Damour,
Deruelle, Ruffini in \cite{damour} and Zouros, Eardley in \cite{zouros}.
These make the existence of exponentially growing modes for the massive 
Klein-Gordon equation very plausible. They found approximate unstable  
modes in the superradient regime, i.e., with frequencies $\omega$ 
satisfying $Re(\omega) < ma/(2Mr_{+})$. These modes become stable
when this condition is violated. The approximations made in these 
papers lead to further restrictions. It turns out that the assumption of, 
both, these restrictions and the bound (\ref{massineq}) derived here
is {\it incompatible} with the assumption of superradience.
 Hence the stability result here does not contradict the
results in these papers, but is complementary instead. 
\footnote{The author is very grateful to J. L. Friedman for directing his
attention to this fact.}
Moreover it suggests that the negation of 
an inequality of the form of (\ref{massineq}) (or some equivalent form) 
{\it is} the superradient condition. For this it should be noted 
that with some effort and along the lines of this paper
it may be possible to improve (\ref{massineq}), i.e., to decrease the
bound. For this the Poincare-Sobolev inequalities mentioned 
at the end of Section 3 should be helpful.  
\newline
\linebreak
{\bfseries Acknowledgements.} 
The author is grateful to B. G. Schmidt for pointing his 
attention to the problem of defining the quasi-normal frequencies 
of the Kerr black hole as resonances and to J. L. Friedman, B. G. Schmidt and 
B. F. Whiting for valuable discussions.

\section{Appendix}

The following theorem used in Section 5 was first proved by 
Dunkel in \cite{dunkel} (compare also \cite{levinson,bellman,hille}).
\begin{theorem}
Let $n \in {\mathbb{N}} \setminus \{0\}$, $a \in {\mathbb{R}}$,
$I := [a,\infty)$ and $I_{0} := (a,\infty)$. 
In addition let $A_0$ be a diagonalizable  complex  $n \times n$ 
matrix and $e^{\prime}_{1} , \dots,   e^{\prime}_{n}$
be a basis of ${\mathbb{C}}^{n}$ consisting of eigenvectors of  $A_0$. 
Further, for each $j \in \{1,\cdots,n\}$ let $\lambda_{j}$ 
be the eigenvalue corresponding to  $e^{\prime}_{j}$ and $P_{j}$ 
be the matrix representing the projection of ${\mathbb{C}}^{n}$
onto ${\mathbb{C}}.e^{\prime}_{j}$ with respect to the canonical basis of 
${\mathbb{C}}^{n}$.
Finally, let $A_{1}$ be a continuous map from $I$ into the complex 
$n \times n$ matrices $M(n \times n,{\mathbb{C}})$ 
such that $A_{1jk}$ is Lebesgue integrable for each 
$j,k \in {1,..., n}$. 
\newline
\linebreak
Then there is a $C^{1}$ map $R:I_{0} \rightarrow M(n \times n,{\mathbb{C}})$
with $lim_{t \rightarrow \infty}R_{jk}(t) = 0$ for each $j,k \in {1,\dots,n}$ 
and such that $u:I_{0} \rightarrow M(n \times n,{\mathbb{C}})$ defined by
\begin{equation} 
u(t):=
\sum^{n}_{j=1}exp(\lambda_{j}t)\cdot (E+R(t))\cdot P_{j} 
\end{equation}
for all $t \in I_{0}$
(where $E$ is the $n \times n $ unit matrix),  maps into the 
invertible  $n \times n $ matrices and satisfies
\begin{equation} 
u^{\prime}(t) = \left(A_{0} + A_{1}(t) 
\right) \cdot u(t)
\end{equation}
for all $t \in I_{0}$.
\end{theorem}
This theorem has the following 
\begin{corollary}
Let $n \in {\mathbb{N}}\setminus \{0\}$; $a, t_0 \in {\mathbb{R}}$ with 
$a < t_0$; $\mu \in {\mathbb{N}}$;
$\alpha_{\mu} := 1$ for  $\mu=0$ and
$\alpha_{\mu} := \mu$ for  $\mu \neq 0$. 
In addition let $A_0$ be a diagonalizable  complex  $n \times n$ 
matrix and $e^{\prime}_{1} , \dots,   e^{\prime}_{n}$
be a basis of ${\mathbb{C}}^{n}$ consisting of eigenvectors of  $A_0$. 
Further, for each $j \in \{1,\cdots,n\}$ let $\lambda_{j}$ 
be the eigenvalue corresponding to  $e^{\prime}_{j}$ and $P_{j}$ 
be the matrix representing the projection of ${\mathbb{C}}^{n}$
onto ${\mathbb{C}}.e^{\prime}_{j}$ with respect to the canonical basis of 
${\mathbb{C}}^{n}$.
Finally, let $A_{1}$ be a continuous map from $(a, t_0)$ into the complex 
$n \times n$ matrices $M(n \times n,{\mathbb{C}})$ for which there is 
a number $c \in (a,t_0)$ such that the restriction 
of $A_{1jk}$ to $[c,t_0)$ is Lebesgue integrable for each 
$j,k \in {1,..., n}$. 
\newline
\linebreak
Then there is a $C^{1}$ map $R:(a,t_{0}) \rightarrow M(n \times n,{\mathbb{C}})$
with $lim_{t \rightarrow 0}R_{jk}(t) = 0$ for each $j,k \in {1,\dots,n}$ 
and such that $u:(a,t_{0}) \rightarrow M(n \times n,{\mathbb{C}})$ defined by
\begin{eqnarray} \label{u}
&u(t):= \cr
&\left\{
 \begin{array}{ll}
  \sum^{n}_{j=1}(t_0-t)^{-\lambda_{j}}\cdot (E+R(t))\cdot P_{j} 
  \ \mbox{for $\mu=0$} \cr
  \sum^{n}_{j=1}exp(\lambda_{j}(t_0-t)^{-\mu})\cdot (E+R(t))\cdot P_{j} 
   \ \mbox{for $\mu \neq 0$ }
 \end{array} \right. \,
\end{eqnarray}
for all $t \in (a,t_{0}) $
(where $E$ is the $n \times n $ unit matrix),  maps into the 
invertible  $n \times n $ matrices and satisfies
\begin{equation} \label{asymptoticofu}
u^{\prime}(t) = \left( \frac{\alpha_{\mu}}{(t_0-t)^{\mu+1}}A_{0} + A_{1}(t) 
\right) \cdot u(t)
\end{equation}
for each $t \in (a,t_{0})$.
\end{corollary}

\end{document}